\documentclass{optica-article}

\journal{opticajournal} 

\articletype{Research Article}

\usepackage{lineno}
\usepackage{physics}
\usepackage{amsmath}
\usepackage{pifont}
\usepackage{enumitem}
\usepackage{mathrsfs}

\begin{document}

\title{Time delay of mean-field interaction in thermal Rydberg atomic gases}

\author{Yuzhuo Wang\authormark{1,2}, Tianxiong Gao\authormark{1,2}, Yufan Niu\authormark{3}, Ying Hu\authormark{1,2}, Linjie Zhang\authormark{1,2}, Suotang Jia\authormark{1,2}, Mingyong Jing\authormark{1,2,\dag}, Yanhong Xiao\authormark{1,2,3,*}}

\address{\authormark{1} State Key Laboratory of Quantum Optics Technologies and Devices, Institute of Laser Spectroscopy, Shanxi University, Taiyuan, Shanxi 030006, China.\\
\authormark{2}Collaborative Innovation Center of Extreme Optics, Shanxi University, Taiyuan 030006, China.\\
\authormark{3} Department of Physics, State Key Laboratory of Surface Physics and Key Laboratory of Micro and Nano Photonic Structures (Ministry of Education), Fudan University, Shanghai 200433, China.}
\email{\authormark{\dag}jmy@sxu.edu.cn} 
\email{\authormark{*}yxiao@sxu.edu.cn} 

\begin{abstract} 
	Mean-field theory is commonly employed to study nonequilibrium dynamics in hot Rydberg atomic ensembles, but the fundamental mechanism behind the generation of the mean-field interactions remains poorly understood. In this work, we experimentally observe a time-delay effect in the buildup of mean-field interaction, which reveals the key role of collision ionization. We analyze the relevant collision channels and propose a microscopic mechanism that quantitatively explains the hysteresis window observed in optical bistability. Then, using square-wave modulation spectroscopy (SMS) to monitor the growth of the mean-field interaction, we experimentally demonstrate a delay in its dynamical buildup following the initial Rydberg excitation. Finally, we demonstrate how this delay effect may help understand the recently observed self-sustained oscillations in a thermal Rydberg gas. Our findings provide compelling evidence for the contribution of ionization processes in the nonequilibrium dynamics of thermal Rydberg gases, a system of growing interest for quantum sensing and quantum information science.
\end{abstract}


\section{Introduction}
Rydberg atomic gases are an important platform for exploring nonequilibrium dynamics~\cite{Charles2013prl3,Charles2018nc,Ding2020PRX}, quantum information science~\cite{Charles2013prl2,Rempe2019np,LiLin2022Nc,ZSL2019nph} and quantum sensing~\cite{Shaffer2012np,Jing2020np,Yuan2023RPP,HB2024T}. The strong interaction between atoms depends on the principal quantum number, introducing high and controllable nonlinearities~\cite{charles2017np,vv2018n,Rempe2019np,LI2014prl,IA2021CP}. The spacing of selectable Rydberg levels ranges from megahertz to terahertz, enabling new external control and detection capabilities~\cite{Charles2013prl2,Charles2018nc,Ding2022np,fan2023OE,RS2024NJP}. Thanks to the controllable interaction even at room temperature, rich phenomena have been observed, such as optical bistability (OB)\cite{Charles2013prl3,Charles2016pra,JG2017jb}, self-organization~\cite{Ding2020PRX} and  self-sustained oscillation (SSO)\cite{Ding2024SciA,YL2023TC,Charles2023prl,Jiao2024TC}. Notable applications have also emerged, including single-photon source\cite{PT2018science} and enhanced metrology at the critical point\cite{Ding2022np,XGY2024SA}. In most of these studies, the relevant many-body physics are treated using mean-field theory.

In the mean-field framework of hot Rydberg atoms, the complicated atomic interactions are often accounted for by replacing the excitation laser's bare detuning, $\Delta$, with an effective, nonlinear detuning, $\Delta-\Delta_{\text{eff}}$, where $\Delta_{\text{eff}}$ is a function of the Rydberg population $\rho_{\text{rr}}$. Common phenomenological expressions, such as $\Delta_{\text{eff}}=V\rho_{\text{rr}}^\beta$ or $\Delta_{\text{eff}}=\sum_{\beta=1}^n V_{\beta}\rho_{\text{rr}}^\beta$, are typically adopted~\cite{Charles2013prl3,Charles2023prl,Ding2022np,YL2023TC}, where $V$ and $V_{\beta}$ represent the interaction strength, and $\beta$ is the power coefficient of $\rho_{\text{rr}}$ depending on the function for the interaction. While this mean-field formalism has been useful in explaining many experimental observations, including those in spectroscopy and nonlinear phenomena~\cite{Charles2013prl,Ding2022np}, the microscopic mechanism  by which the Rydberg population contributes to the mean-field interaction remains poorly understood. 

For instance, in the case of the OB spectrum, Doppler-induced frequency shifts would typically cause atoms in different velocity groups to undergo different detunings, potentially smearing the phase transition points in the OB spectra~\cite{WJH2019JB}. However, OB windows and sharp transitions are routinely observed experimentally, suggesting that the phase transition is driven by a global change in the atomic state, rather than by the detuning of the individual velocity classes. Moreover, in case of the Rydberg SSO,  the effective detuning, $\Delta_{\text{eff}}(\rho_{\text{rr}})$, is thought to play a key role in driving the Hopf bifurcation, yet the underlying microscopic mechanism is still debated due to quantitative discrepancies between the experimentally observed oscillation frequencies and the theory~\cite{Charles2023prl,Jiao2024TC}. Furthermore,  SSO is observed even in the absence of a cavity or direct atom-atom interactions when delayed feedback is introduced~\cite{LGB2024arxiv}. This suggests that delayed feedback may play a fundamental role in generating these oscillations. In this context, delayed feedback refers to a process where a signal from the system acts back on the system after a significant period of time has elapsed.

Several studies have suggested that the mean-field effect in a Rydberg gas may originate from the ionization process~\cite{Haroche1999JB, PT2013prl, Harald2016pra, Harald2019pra}, indicating that the Rydberg population does not contribute to the mean-field interaction immediately. It has also been shown that the OB hysteresis results from an avalanche ionization process~\cite{Harald2016pra}. While this ionization-to-plasma transition has been experimentally examined using electrical readout~\cite{PT2013prl}, external electric fields can alter plasma dynamics. Alternatively, plasma properties have been inferred by comparing experimental spectra with numerical models~\cite{Harald2019pra}, but this approach is indirect and complex. Additionally, a ``forest fire'' model has been proposed to describe the mean-field effect~\cite{Ding2020PRX}, accounting for the threshold effect and enhanced dephasing, yet the underlying microscopic mechanism remains unclear.

Here, we experimentally demonstrate and theoretically analyze a time delay effect in the buildup of mean-field interaction, highlighting the importance of microscopic process in understanding the nonequilibrium dynamics in a Rydberg atomic gas. We begin by describing an ionization model to account for the plasma-induced mean-field interaction. We show that the lower branch of the OB corresponds essentially to the single-atom spectrum, where the mean-field interaction is only perturbative. Upon reaching the collision ionization threshold, the system exhibits a ``jump'' to the upper branch. We analyze the consequence of avalanche process and threshold effect in the OB spectrum. Next, using the SMS method, we experimentally observe the delayed dynamical buildup of mean-field interaction after the Rydberg population is created. This provides new insights into the origin of the mean-field interaction and helps clarify the dynamics in bistable systems. Finally, we apply our model to analyze SSO. Our findings shed light on the fundamental mechanisms of mean-field interactions in hot Rydberg atoms and offer valuable insights into nonequilibrium physics in these systems. Moreover, the results and the SMS method we propose have broad applicability for investigating many-body physics and nonlinear optical devices based on atomic systems.

\section{Theoretical model}

\subsection{Model on the coupling between atoms and plasma}

\begin{figure}[t]
	\centering
    \includegraphics[width=0.88\textwidth]{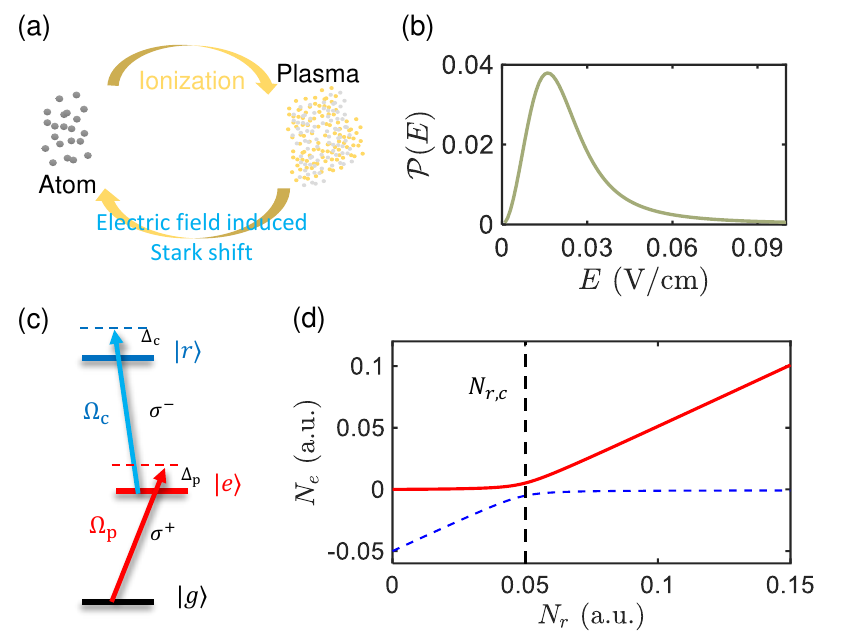}
	\caption{ Mechanism of mean-field interaction in thermal Rydberg gases. (a) A schematic diagram of atom-plasma coupling. (b) An example of $\mathcal{P}(E)$ with $E_c=0.01$~V/cm. (c) Two-photon excitation diagram. (d) Solutions of $N_e$ when scanning $N_r$ at steady state, where red solid and blue dashed lines are positive and negative solutions respectively. The threshold $N_\text{r,c}$ is marked as a black dashed line. The parameters used in the simulation were $a_1=1\times 10^{-5}$, $a_2=1\times 10^{-4}$, $a_3=0.02$, $\Gamma_e=0.001$, $\Gamma_e'=0.02$.}
	\label{fig:th}
\end{figure}

Since the ionization process of the Rydberg state does not retain phase information, the plasma produced has no coherence with the atoms. Instead of a 4-level model which includes plasma as a state\cite{Harald2019pra}, we simplify the system as a 3-level atom whose energy level is shifted by the electric field due to the plasma, as illustrated in Fig.\ref{fig:th}(a). The DC Stark shift from an electric field $E$ is
\begin{equation}
	\Delta_s=-\frac{1}{2} \alpha| E|^2
		\label{eq:DE}
\end{equation}
 where $\alpha$ is the atomic polarizability of Rydberg state, calculated by the software package ARC~\cite{arc2017}. While in a plasma, the electric field distribution is described by Holtsmark function\cite{Holtsmark1919,Harald2016pra,Anderson2017}:
\begin{equation}
	\mathcal{P}(E) = \mathcal{H}(E/E_c)/E_c 
		\label{eq:PE}
\end{equation}
\begin{equation}
	\mathcal{H}(\beta)= \frac{2}{\pi\beta} \int^\infty_0  x \sin x e^{-(x/\beta)^{3/2}} \dd x,
		\label{eq:PE2}
\end{equation}
where $E_c$ is a characteristic electric field from the charged particle density $N_e$, 
\begin{equation}
	E_c= \frac{|e|}{2\epsilon_0} (\frac{4}{15}N_e)^{2/3}
		\label{eq:Ec}
\end{equation}
with electron charge $e$ and vacuum permittivity $\epsilon_0$. An example of $\mathcal{P}(E)$ with $E_c=0.01$~V/cm is shown in Fig.\ref{fig:th}(b). This distribution introduces two effects on the Rydberg state, an effective frequency shift $\Delta_\text{eff}$ 
\begin{equation}
\Delta_{\text{eff}}=\int^\infty_0 \dd E  \mathcal{P}(E) \Delta_s(E)
		 \label{eq:D_eff}
 \end{equation}
and a dephasing
\begin{equation}
   \hat L_r=\sqrt{\Gamma_s}|r\rangle \langle r|.
		\label{eq:R phase decay}
\end{equation}
$\hat L_r$ reflects the dephasing process which broadens the atomic spectrum but does not change the population directly. We cut off the integration in Eq.(\ref{eq:D_eff}) at $10E_c$ in the numerical calculation. The Hamiltonian $\hat H_0$ including the DC Stark shift of Rydberg levels has the form  (we take $\hbar=1$)
\begin{equation}
	\hat H=-\frac{1}{2}\left[
		\begin{array}{ccc}
		  0        & \Omega_p^*   & 0            \\
		  \Omega_p & 2\Delta_p    & \Omega_c^*    \\
		  0        & \Omega_c     &  2(\Delta_p+\Delta_c+\Delta_{\text{eff}} )\\
		\end{array}
		\right],
		\label{eq:MH}
\end{equation}
where $\Omega_{p,c}$ and $\Delta_{p,c}$ are Rabi frequencies and detunings for the two relevant transitions, as in Fig.\ref{fig:th}(c). The system's density matrix evolves under a Lindblad master equation
\begin{equation}
	\frac{\dd \hat \rho(t)}{\dd t}=-i[\hat H, \hat \rho(t)]+\sum_k (\hat L_k\rho\hat L_k^+ -\frac{1}{2}\{\hat \rho(t),\hat L_k^+ \hat L_k   \}),
		\label{eq:MasterA}
\end{equation}
where $\hat L_k (k=1,2,3,r)$ represents the collapse operators for decay channels including
\begin{equation}
	\hat L_1=\sqrt{\Gamma_\text{eg}}|g\rangle \langle e|, \quad
	 \hat L_2=\sqrt{\Gamma_\text{rg}}|g\rangle \langle r|, \quad
	\hat L_3=\sqrt{\Gamma_\text{re}}|e\rangle \langle r|
		\label{eq:L}
\end{equation}
and the dephasing $\hat L_r$ in Eq.(\ref{eq:R phase decay}). Up to now, we have connected the plasma to Rydberg level shift. The electric field from the plasma not only shifts the Rydberg level, but also introduces an extra dephasing, which contributes to the systems' dynamics described with 
Eq.(\ref{eq:MasterA}). Next, we will discuss the plasma generation. Considering that the charged particle production corresponds to Rydberg atom loss, we include the loss in $\hat L_2$ to ensure population conservation. It is a good approximation to treat the ground state atoms as a reservoir, since in hot atomic ensemble the Rydberg atom number that can be excited is much less than ground state atoms due to the wide velocity distribution and the associated Doppler shift.

\subsection{Collision ionization and threshold effect}\label{sec:ionization}

The charged particles in plasma are from collision ionization. According to previous studies on Rydberg ionization\cite{Haroche1999JB,LWH2005prl,Harald2019pra}, there can be four collision ionization channels between:
\begin{enumerate} [itemsep=2pt,topsep=0pt,parsep=0pt]
	\item[\ding{172}] Rydberg and ground state atoms,
	\item[\ding{173}] Rydberg and Rydberg state atoms,
	\item[\ding{174}] Rydberg atom and electrons,
	\item[\ding{175}] Rydberg atom and Rydberg ions.
\end{enumerate}	
The ionization rate is determined by the particle density, particle speed and collision cross-section. Electrons and Rydberg ions have similar density $N_e\approx N_\text{ion}$, so the effects of \ding{174} and \ding{175} can be merged into one term mathematically. 
Based on the ionization mechanism discussed above, a rate equation of electron density $N_e$ can be derived, which was also mentioned in ref\cite{Harald2019pra}:
\begin{equation}
	\frac{\dd N_e}{\dd t}=a_1 N_g N_r+a_2 N_r^2+a_3 N_e N_r-\Gamma_e N_e-\Gamma_e' N_e^2.
		\label{eq:Nc}
\end{equation}
Here $N_g$ and $N_r$ are atom densities for the ground state and Rydberg state respectively. We normalize the densities by setting atomic density $N_a=1$, and thus $N_r, N_e \ll N_g$ and they are all dimensionless. The first three terms on the right hand side of Eq.(\ref{eq:Nc}) stand for \ding{172}, \ding{173} and \ding{174}+\ding{175} , respectively. Coefficients $a_1$, $a_2$ and $a_3$  are constants related to the collision cross-section and the particles' most probable speed. Considering the cross-sections relation \ding{174}$\approx$ \ding{173}$>$ \ding{172}\cite{Haroche1999JB}, and electrons have a much larger probable speed, we have $a_3\gg a_2>a_1$. $\Gamma_e$ is the single-body decay rate of charged particles, which is influenced by multiple factors such as transit effect, and the last term $\Gamma_e' N_e^2$ is the electron loss due to electron-ion recombination. The constants $a_1, a_2, a_3, \Gamma_e$ and $\Gamma_e'$ are all frequency quantities, and we set their units as the decay rate $\Gamma_\text{eg}$ (see Eq. (\ref{eq:L}))in numerical works.

Eq.(\ref{eq:Nc}) describes an avalanche process: ionization generates charged particles, giving rise to a larger Rydberg-electron collision probability. Eventually, a steady state is reached when the ionization rate is limited by the Rydberg population loss and electron-ion recombination loss. This avalanche process could be the source of the forest-fire model in \cite{Ding2020PRX}, where it was shown that the interaction and relevant dephasing appear when the Rydberg density exceeds a threshold $N_\text{r,c}$. This threshold can be reached by adjusting the detuning or the laser Rabi frequency, and it is closely related to the transmission jump in bistability spectra. Here we emphasize that the density threshold corresponds to a collision threshold, as demonstrated in cold atom ensembles\cite{Haroche1999JB,PP2000prl,wm2013prl}. The mechanism and expression of the avalanche threshold can be derived through an analysis of Eq (\ref{eq:Nc}). For an avalanche to take place, the term $a_3 N_e N_r$ must surpass the combined contributions from other ionization processes $(a_3 N_e N_r>a_1 N_g N_r+a_2 N_r^2)$ as well as the ion dissipation rate $(a_3 N_e N_r>\Gamma_e N_e+\Gamma_e' N_e^2)$. When the ion population is small, the dissipation term is primarily governed by $\Gamma_e N_e$. Thus, the condition for avalanche is $a_3 N_e N_r>\Gamma_e N_e$, leading to the threshold $N_\text{r,c}= \Gamma_e/a_3$.  This is consistent with the vertex point of the steady-state solution 
\begin{equation}
N_\text{r,c}=\frac{\Gamma_e-2\Gamma_e'a_1/a_3}{a_3+4\Gamma_e'a_2/a_3}\approx \frac{\Gamma_e}{a_3}\label{eq:Nrc}
\end{equation}
when the approximation above holds. We note that the quadratic equation $\dd N_e/\dd t=0$ has two solutions for $N_e$ as shown in Fig.\ref{fig:th}(d). The system follows the positive solution (red line) because the negative is non-physical, while $N_\text{r,c}$ has been marked in the figure.  As the avalanche progresses and the ion population grows rapidly, the dissipation process becomes dominated by $\Gamma_e' N_e^2$. At this stage, the avalanche ceases, and the system reaches a steady state, where $\dd N_e/ \dd t=a_3 N_e N_r-\Gamma_e' N_e^2$. In this steady state, the ion density $N_e$ is directly proportional to $N_r$ and our model reduces to $V\rho_\text{rr}^\beta$.
 
Although channels \ding{172} and \ding{173} contribute minorly to ionization rates, the electrons generated from these channels serve as the initial seeds required to trigger other collision processes. The above discussion provides an intuitive understanding of the avalanche process and its threshold. However, a more rigorous analysis should refer to the complete model, which combines Eq.(\ref{eq:Nc}) and Eq.(\ref{eq:MasterA}).


\section{Experimental demonstration}\label{Sec:exp}

\subsection{Experiment setup}\label{Sec:setup}

To experimentally study the mean-field interaction, we adopt a two photon excitation scheme. 
We perform electromagnetically induced transparency (EIT) spectroscopy in an enriched $^{87}$Rb vapor cell (cylindrical) with length of 7.5 cm and diameter of 2.5 cm. As shown in Fig.\ref{fig:th}(c), a probe (red arrow, wavelength 780 nm, Rabi frequency $\Omega_p$) and a coupling (blue arrow, 480 nm, $\Omega_c$) laser field excite atoms from the ground state $|g\rangle$ to a Rydberg state $|r\rangle$ via an intermediate state $|e\rangle$.  The frequencies of both lasers are calibrated by the Rydberg EIT spectrum at room temperature and at relatively low laser power where the plasma effect is negligible, and are locked to a cavity with a finesse of about 10000.  $\lambda/2$ and  $\lambda/4$ waveplates are used for both fields to compensate for the polarization distortion by the dichroic mirrors. A differential detection~\cite{Ding2020PRX,Jing2020np,YL2023TC} by a balanced photo-detector is employed between the probe beam in the EIT and a spatially separated reference probe beam in the same vapor cell. By controlling the power of the heating layer, the temperature of the vapor cell can be varied from 23$^\circ $C to 81$^\circ $C.  The relevant energy levels are $|g\rangle=|5S_{1/2},F=2\rangle$,  $|e\rangle=|5P_{3/2},F=3\rangle$, and $|r\rangle=|57S_{1/2}\rangle$. The probe and coupling lasers' polarizations are $\sigma^+$ and $\sigma^-$ respectively to drive the selected transitions, with beam waists of 0.30 mm and 0.55 mm respectively. 

We keep $\Delta_p=0$ and scan $\Delta_c$ when recording the EIT spectra. In addition, we study the dynamical Rydberg-excitation process to reveal the coupling dynamics between atoms and the plasma. We keep the red laser on, and modulate the blue laser with an acoustic optical modulator, driven by a square-wave amplitude modulated radio-field (RF). The rising edge of the pulse from the modulation is less than 0.2~$\mu$s, characterized with a photo-detector. The growth of the Rydberg population can be seen from the red laser's transmission, allowing for the study of the ionization process through the rising or falling edges of the transmission. Information of nonequilibrium can be extracted with such a method, which we name as square-wave modulation spectroscopy (SMS). 

\begin{figure}[t]
	\centering
    \includegraphics[width=0.81\textwidth]{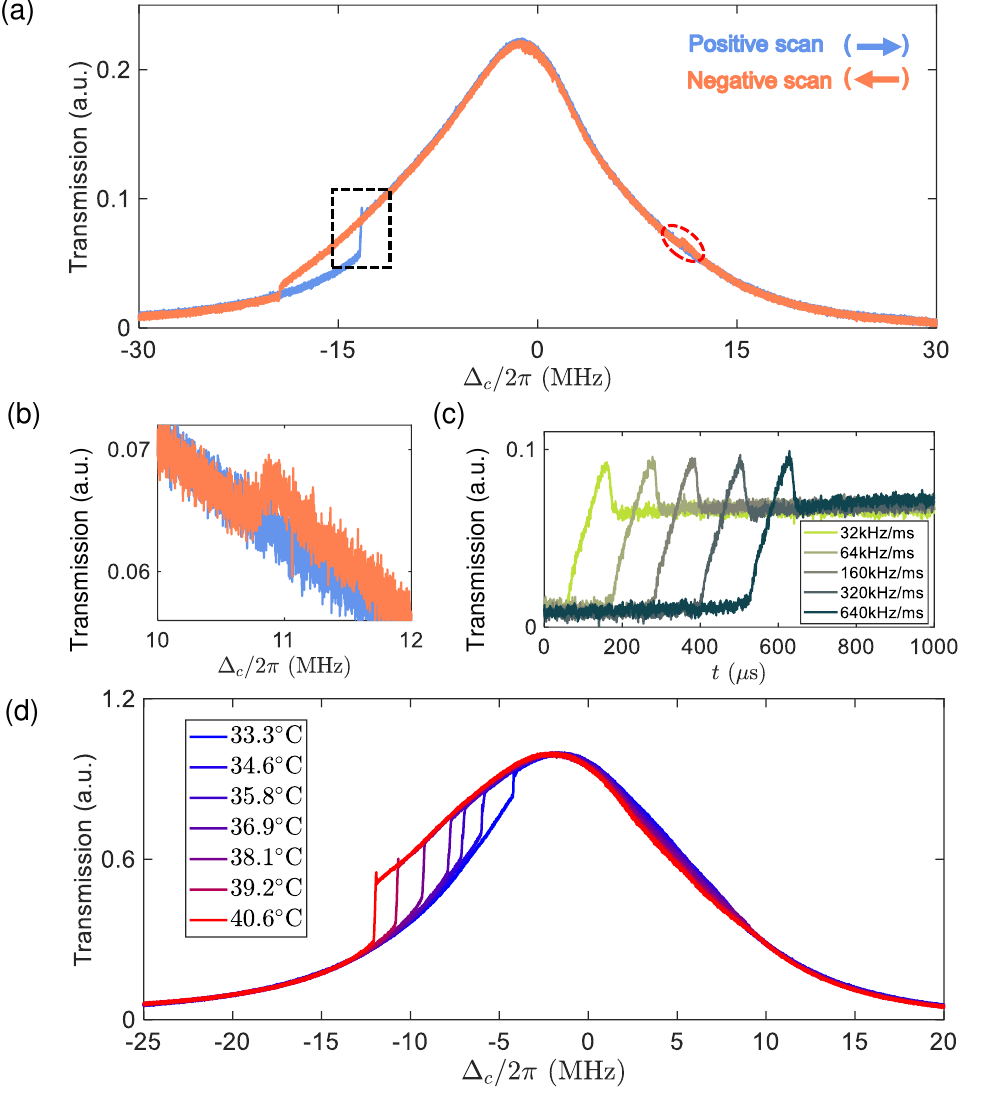}
	\caption{
		OB spectra analysis. (a) A typical OB transmission at $T_\text{cell}=47.9~^\circ$C and $\Omega_p/2\pi=35.8$ MHz, $\Omega_c/2\pi=1.1$ MHz.  (b) The jump at blue detuning of an OB spectrum, which is a zoom-in of red dashed circle in (a). (c) Spectra across the sharp edge, marked as the dashed black box in (a), under different frequency scan speed. Each spectrum is displaced with an offset time for viewing purpose. The parameters are the same as (a). (d) Transmission spectra under different cell temperature with $\Omega_p/2\pi=24.8$ MHz, $\Omega_c/2\pi=1.6$ MHz. }
	\label{fig:OB}
\end{figure}

\subsection{OB hysteresis and sharp edge}\label{sec:OBjump}

\begin{figure}[t]
    \centering
    \includegraphics[width=0.95\textwidth]{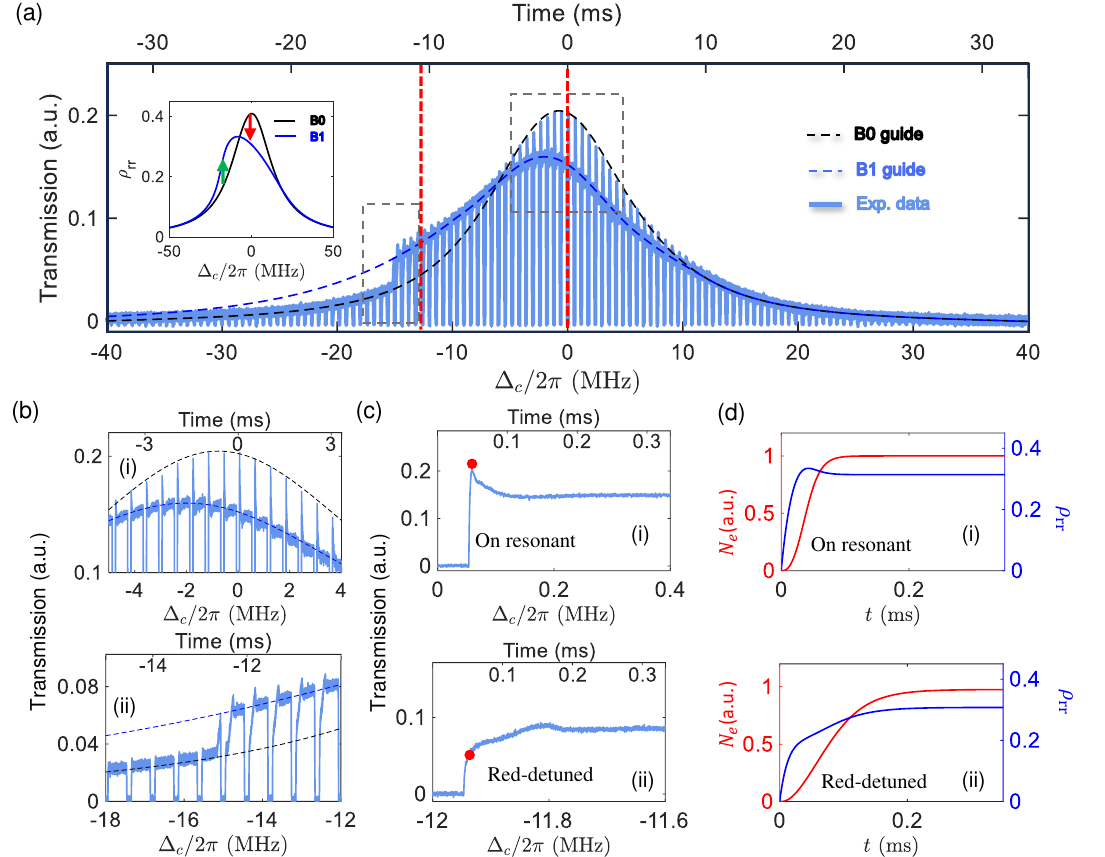}
        \caption{ Time-delay of mean-field interaction. (a) Transmission with the 2 kHz modulated blue laser at $T_\text{cell}=47.5~^\circ$C and $\Omega_p/2\pi=30.8$ MHz, $\Omega_c/2\pi=1.3$ MHz. Time axis is also provided and $t=0$ is aligned with the on resonant point. Black solid line is a guideline to represent a single-atom EIT spectrum. The inset is calculated curves with B0 (black) and B1 (blue) branches. Arrows show the relaxation direction of the transmission.
	    (b,c) Zoom-in figures of (a) at near resonant and red-detuned regions under different magnification, marked with grey dashed boxes (b) and red dashed lines (c) in (a). In (c) $t=0$ is adjusted to the start of this figure for viewing purpose. (d) Numerically calculated time dependent $\rho_\text{rr}$ and $N_e$. The parameters were set to $\Omega_p/2\pi=15$ MHz, $\Omega_c/2\pi=1.2$ MHz, $a_1=3\times 10^{-5}$, $a_2=0.01$, $a_3=0.02$, $\Gamma_e=0.02$, $\Gamma_e'=0.02$.}\label{fig:spect}
\end{figure}

First, we revisit the OB hysteresis in a hot Rydberg gas using our model. In principle, the OB window should be seen when $N_r$ crosses the threshold, both on the red and blue detuning sides of the EIT spectrum.  In Fig.\ref{fig:OB}(a), we show an OB transmission spectrum with $\Omega_p/2\pi=35.8$ MHz, $\Omega_c/2\pi=1.1$ MHz and Rb cell temperature $T_\text{cell}=47.9^\circ $C. The OB window on the blue detuning side (Fig.\ref{fig:OB}(b)) is relatively small, which can be explained in the following. Consider the steady state equilibrium spectrum with and without plasma-mediated interaction as B1 and B0, as calculated using our model and shown in Fig.\ref{fig:spect}(a) inset, where B1 is shifted to the left relative to B0 due to the DC Stark shift (see Eq.(\ref{eq:DE})). We note that, as indicated in Fig.\ref{fig:th}(b) and also provable by our calculation, the average E field (near the peak location) is comparable to the width of the distribution curve, which suggests the DC stark shift $\Delta_\text{eff}$ and the inhomogeneous electric field broadening are of the same scale~\cite{Harald2019pra}. This leads to the fact that B0 and B1 branches are closer when $\Delta_c>0$ and are more apart giving a larger OB window when $\Delta_c<0$ (see Fig.\ref{fig:spect}(a) inset). The OB window for blue detuning would become more significant if the difference between two branches were larger at the $N_r$ threshold location.  

One might consider whether the sharp edge of an OB window is detuning dependent and benefits the precision frequency measurement\cite{Ding2022np}. To understand better the OB property and its indication to metrology applications, we scan $\Delta_c$ across the sharp edge (dashed black box in Fig.\ref{fig:OB}(a)) under different speed, and the results are shown in Fig.\ref{fig:OB}(c). Lines with deeper color stand for a faster scanning, and a time offset is added to different lines in plotting for viewing purpose. The rising edges are nearly identical for all scanning speeds, illustrating that the rise in the transmission does not depend on the amount of detuning change, but is only determined by the location of the threshold. In other words, the system's dynamics is dominated by the avalanche process, which is triggered when the detuning approaches the threshold. From our findings, we conclude that this detuning-decoupled sharp edge does not directly improve the precision of frequency measurement. However, its ability to assist in locating the jump point with greater sensitivity could be advantageous for metrology applications.

Furthermore, we study the jumps under different Rb cell temperature. The results for positive scanning are in Fig.\ref{fig:OB}(d), where the transmission is normalized to compensate for the changing optical depth. Spectra almost overlap at the start, and begin to diverge at the jumps, where the mean-field interaction begins to play a role. As the Rb cell temperature increases, the atomic density $N_a$ is higher, and the constant $a_3$ in Eq.(\ref{eq:Nc}) has negligible change, so a lower critical Rydberg population $\rho_\text{rr,c}=N_\text{r,c}/N_a$ is required with the same $N_\text{r,c}$. This is consistent with our experiment observations: at higher atomic density , the jump appears at a larger red detuning corresponding to a lower Rydberg population. 


\subsection{Observation of delayed mean-field interaction}\label{sec:OBswitch}

Then we report experiment evidence of the time delay in mean-field interaction upon Rydberg excitation. We modulate the blue laser at 2 kHz and with a square-wave of 80$\%$ duty cycle. The Rabi frequencies of the red and blue lasers are $\Omega_p/2\pi=30.8$ MHz and $\Omega_c/2\pi=1.3$ MHz respectively, calculated from the input powers and beam waists~\cite{arc2017}. The blue laser is off for $100~\mu$s in each modulation period, which is long enough for the Rydberg population to diffuse out of the excitation area.  We set the red laser on resonant, and scan the blue laser frequency. SMS is shown in Fig.\ref{fig:spect}(a), while for comparison the spectrum without modulation is in Fig.\ref{fig:OB}(a). The Fig.\ref{fig:spect}(b) shows the details of the probe transmission evolution. In both the near two-photon resonance and red-detuned regions, the transmission reaches an initial value (marked by the red dots in the zoom-in pictures in Fig.\ref{fig:spect}(c)), goes through some transits and then reaches a relatively steady level. This steady level is higher (lower) than the initial value in the red-detuned (near resonant) region. Such phenomenon can be explained below, and is evidence for the time delay in the mean-field.    




The dynamics subsequent to the activation of the blue laser can be divided into two distinct stages: first, the system experiences a normal EIT excitation (without plasma-mediated interaction) to branch B0, which is indicated by the dashed black line formed by connecting the red dots; Second, the mean-field interaction is generated due to collision, and then the system relaxes to a steady state considering plasma effect, named B1 branch which is indicated by the dashed blue line formed by connecting the steady transmission level in each modulation period. The calculated spectra of B0 and B1 (Fig.\ref{fig:spect}(a) inset) using our model can be used to illustrate this process: the system goes to the black curve B0 first and then relaxes to the blue curve B1, as indicated by the arrows. In the red-detuning (resonant) region, the green (red) arrow shows increase (decrease) of probe transmission, which agrees with the experiment trends in Fig.\ref{fig:spect}(c).

To verify the delay effect introduced by our model, we conducted simulation of the dynamics for $\rho_\text{rr}$ and $N_e$ when driving the atoms from the ground state. The time dependence curve is presented in Fig.\ref{fig:spect}(d), and the trends of $\rho_\text{rr}$ curves  qualitatively agree with the measured transmission in Fig.\ref{fig:spect}(c). The mean-field interaction doesn't show a significant effect until the avalanche process begins and $N_e$ grows fast. At this moment, the system enters its second-step dynamics. Using such a delay effect, we have obtained profiles of non-interaction branch through our modulation method even beyond the OB windows. Overall, the above experiments and simulation suggest that, the mean-field interaction is delayed after the Rydberg population is created.

\subsection{Time dependent energy shift from nonequilibrium spectrum}\label{sec:NS}

\begin{figure}[t]
    \centering
    \includegraphics[width=0.95\textwidth]{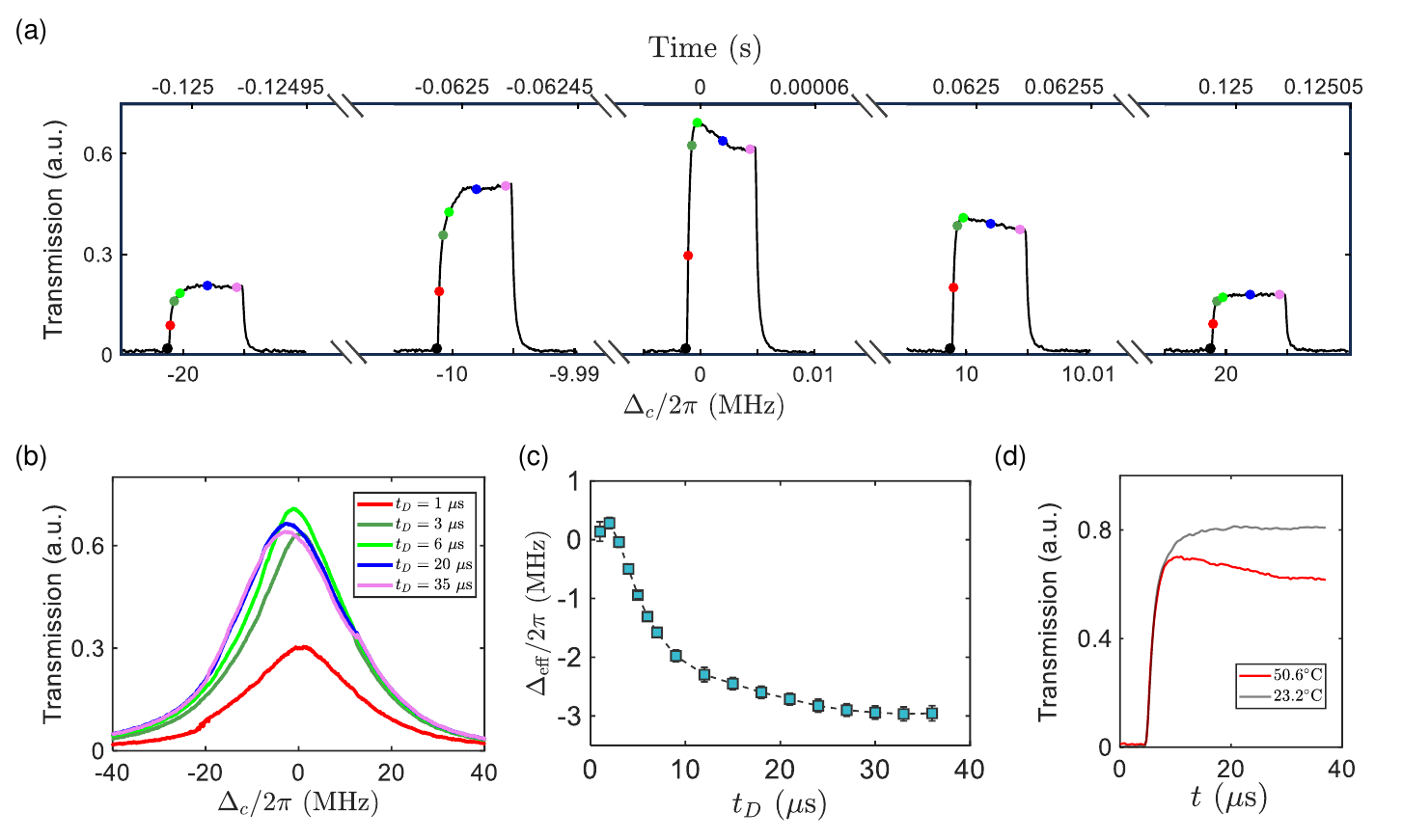}
        \caption{ Dynamical buildup of mean-field interaction. (a) An example of SMS sampling. Black dots mark the different switching-on time $\{t(\Delta_c)\}$, and dots in other colors are sample points at $t(\Delta_c)+t_D=1,3,6,20,35~\mu$s. The transmission pulses are shown around $\Delta_c/2\pi=-20,-10,0,10,20$~MHz.   Other parameters are $T_\text{cell}=50.6^\circ$C, $\Omega_p/2\pi=47.5$ MHz, $\Omega_c/2\pi=1.2$ MHz. (b) A global view of sampled SMS at $t_D=1,3,6,20,35~\mu$s. The color of each curve matches that of the dots in (a) for the corresponding  $t_D$. (c) Spectrum shift $\Delta_\text{eff}$ derived from (b). Error bar is the standard deviation of 10 repetitions of the SMS experiments. (d) Rising edges of the probe transmission at high and low atomic densities. The low-density signal is acquired by setting $T_\text{cell}=23.2^\circ$C while keeping other parameters unchanged. Due to the signal increase caused by the reduction of the optical depth, a rescaling factor on the voltage is used for easier comparison. }\label{fig:SMS}
\end{figure}

We confirmed above that the mean-field interaction takes effect within a finite delay time after the initial Rydberg population generation, instead of being instantaneous. In this section we show that this delay time of the mean-field interaction, as well as the time evolution of the Stark shift, can both be extracted from SMS. We reduce the scanning speed to 50 kHz/ms, and modulate the blue laser at 10 kHz with a square-wave of 40$\%$ duty cycle, which gives the laser's on and off time in each cycle 40$~\mu$s and 60$~\mu$s respectively.  We denote the switching-on time in each cycle as $\{t(\Delta_c)\}$ as marked by the black spots in Fig.\ref{fig:SMS}(a), with $\Omega_p/2\pi=47.5$ MHz, $\Omega_c/2\pi=1.2$ MHz. The transmission points collected at $\{t(\Delta_c)+t_D\}$ of all cycles are plotted and form a new spectrum $T(t_D,\Delta_c)$, which we call sampled-SMS. Here $T(t_D,\Delta_c)$ is a nonequilibrium spectrum at delay time $t_D$ after the blue laser is on. There are 62.5 sampled points within 1 MHz detuning range, which ensures a relatively smooth spectrum. We can derive the interaction induced shift based on $T(t_D,\Delta_c)$, which includes the interaction accumulated from 0 to $t_D$. Sampled-SMS $T(t_D,\Delta_c)$ with $t_D=1,3,6,20,35~\mu$s are shown in Fig.\ref{fig:SMS}(b). The curve color matches that of the dots in (a) for the corresponding $t_D$. $\Delta_\text{eff}(t_D)$ is represented by the detuning at the peak of the curve $T(t_D,\Delta_c)$. It gradually moves to the red side, indicating the growth of the mean-field interaction.

Since the sampled SMS are neither Gaussian nor a Lorentzian shape, we obtain the peak location through a weighted average:
\begin{equation}
	\Delta_\text{eff}(t_D)=\frac{\Delta_c\times T(\Delta_c,t_D)}{\sum T(\Delta_c)}
		\label{eq:maxT}
\end{equation}
This is performed within the region where $T(t_D,\Delta_c)$ is over its 95$\%$ maximum, and $T(t_D,\Delta_c)$ itself acts as the weight. Results of $\Delta_\text{eff}(t_D)$ are shown in Fig.\ref{fig:SMS}(c). $\Delta_\text{eff}$ goes to the red-detuning direction after a small delay, and stabilizes at approximately $2\pi\times(-3.0)$ MHz by $t_D=30~\mu$s.  This indicates that the system evolves into its steady state, where the condition $\dd N_e/\dd t=0$ is satisfied, as described by Eq. (\ref{eq:Nc}). We note that this time delay is much longer than the relaxation of the Rydberg population excitation, which we obtained from a low-density SMS (Fig.\ref{fig:SMS}(d)). Therefore the gradual increase of $\Delta_\text{eff}(t_D)$ reflects the buildup process of the mean-field interaction, or the process of ionization. 

\begin{figure}[t]
    \centering
    \includegraphics[width=0.9\textwidth]{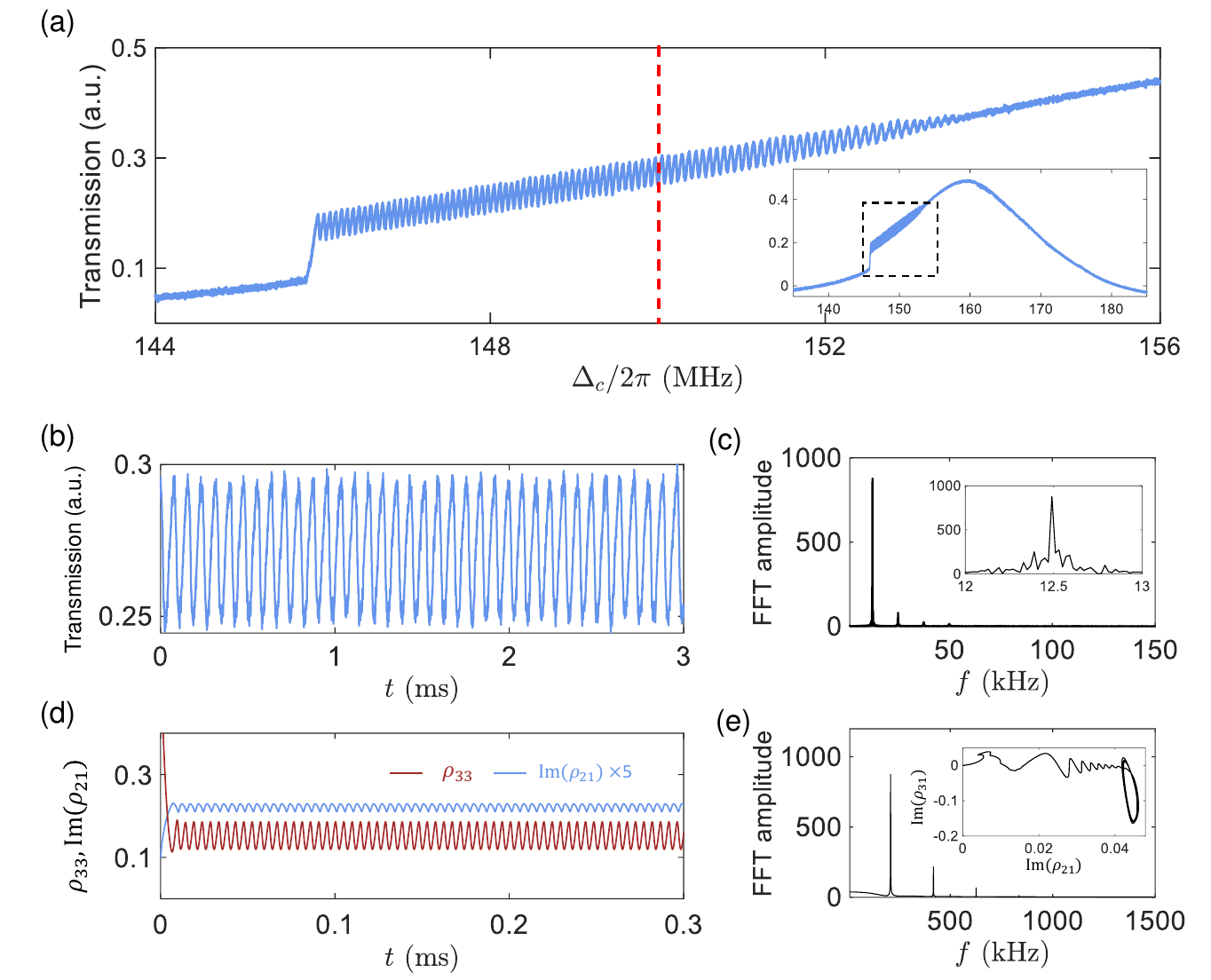}
        \caption{ Delayed mean-field interaction and SSO. (a) A part of SSO spectrum with $\Omega_p/2\pi=36.9$ MHz, $\Omega_c/2\pi=1.3$ MHz and $\Delta_p/2\pi=-100$ MHz. Inset is a global view. (b) The measured time dependent transmission at  $\Delta_c/2\pi=150$ MHz. (c) Frequency spectrum of (b), and the inset is a zoom-in picture around 12.5 kHz. (d) Numerical simulation of time-evolution of $\rho_{33}$ and $\text{Im}(\rho_{21})$, with $\Omega_p/2\pi=24$ MHz, $\Omega_c/2\pi=1.2$ MHz, $\Delta_p/2\pi=-99$ MHz, $\Delta_c/2\pi=163.8$ MHz, $v=95$~m/s. (e) Spectrum of $\text{Im}(\rho_{21})$ in (d). Inset is the trace of coordinate (Im$(\rho_{21})$, Im$(\rho_{31})$).
		 }\label{fig:SSO}
\end{figure}

\subsection{Impact of delayed mean-field interaction in SSO}\label{sec:SSO}

SSO is a typical phenomenon in nonequilibrium systems. Under proper experimental parameters, the value of an observable can display oscillations, as reported in atom-cavity systems \cite{AC1995OC,XM2004praLC,KH2022science}, hot Rydberg atomic gases \cite{YL2023TC,Ding2024SciA,Charles2023prl,Jiao2024TC}, liquid crystal~\cite{SYR1983OL,ZHJ1984,Song1984AO} and many other systems~\cite{IK1979OC,SSO1,SSO2,SSO3,SSO4}. This oscillation indicates a limit cycle solution of a nonlinear system, and is sometimes connected to ``time crystal''\cite{YL2023TC,LGB2024arxiv,PXH2024}. We point out that the delayed interaction discussed in Sec \ref{sec:OBswitch} can be a mechanism to provide delayed feedback and generate SSO: Rydberg atoms are ionized to plasma, and the plasma electric field in turn influences the Rydberg excitation, forming a ``feedback loop'' with a certain timescale. We find SSO near the sharp edge as shown in Fig.\ref{fig:SSO}(a) by carefully adjusting the detuning and power of the lasers. The oscillation begins at the transmission jump, and lasts within a range of $~10$ MHz until its amplitude gradually goes down at  $\Delta_c/2\pi=154$ MHz. To investigate the dynamics under a given set of experimental parameters, we fix the detuning at $\Delta_c/2\pi=150$ MHz to analyze the time-fluctuating and oscillatory transmission (Fig.\ref{fig:SSO}(b)), and the corresponding FFT amplitude spectrum is in Fig.\ref{fig:SSO}(c). The oscillation frequency is about 12.5 kHz, with the FWHM linewidth about 20 Hz (see the inset), showing an ultra-stable oscillation. This oscillation is self-sustained with no observable decay. We note that the oscillation period is on the same time scale with the delay time of the mean-field interaction, which further supports the connection between SSO and the ionization process.

A numerical simulation of  $\rho_{33}$ and $\text{Im}(\rho_{21})$ is shown in Fig.\ref{fig:SSO}(d). The parameters used are similar to the experimental parameters except for the notably smaller $\Omega_p$ which is to take into account of the attenuation of the red laser while propagating along the vapor cell. Considering one dimensional Doppler shift in both laser detunings $\Delta_{p,c}'(v)=\Delta_{p,c}+\vec{k}_{p,c}\vec{v}$, with $\vec v$ the atom velocity and $\vec{k}_{p,c}$ the wave vector of the two lasers. The simulation is performed for a certain velocity group of atoms with $v=95$m/s, so the effective two-photon detuning $\Delta_{p}'+\Delta_{c}'$ is small enough to create Rydberg population for a significant mean-field. The FFT amplitude spectrum is in Fig.\ref{fig:SSO}(e), showing similar narrow peaks at 208.9 kHz and its harmonics. Inset is the trace of coordinate (Im$(\rho_{21})$, Im$(\rho_{31})$), which is a two-dimensional projection of state-space evolution. The trace begins at (0,0) and is finally attracted to a ring, proving that the dynamics is a limit cycle. The oscillation frequency is still one order of magnitude larger than the experiment, but this gap is smaller than the simulation in previous works\cite{Charles2023prl,Ding2024SciA}. The remaining deviation between the simulated oscillation frequency and the observed is mainly due to two aspects: First, the simulation contains only atoms of one velocity. The oscillation will be slower when considering the coupling between different velocity groups which requires much greater computational resources. Second, the propagation effects of lasers along the vapor cell, as well as the charge-particle diffusion are not included, which may introduce extra delay mechanisms. A more complete model will be pursued in the near future.
\section{Conclusion}

In summary, we have demonstrated the dynamical buildup of mean-field interactions through ionization in hot Rydberg atomic ensembles. We first analyze the relevant collision channels and explain the threshold effect, highlight how the sharp transition in the OB spectrum results from an avalanche ionization process, which is largely insensitive to detuning. Using SMS, we observe a delay in the buildup of the mean-field interaction and characterize the gradual development of this effect. Finally, we present a high-Q SSO signal and explain it using the ionization model. In contrast to the dipole-dipole interaction, which dominates the dynamics at $\sim$10~ns timescale\cite{PT2018science,TF2013prlb} and for a Rabi frequency as high as GHz level,  the plasma effect provides a more accurate explanation for most nonlinear phenomena.

A more quantitative analysis, however, requires an accurate estimation of the collision scattering cross-section during ion generation, which poses significant challenges in thermal atomic systems. To address this, we are actively exploring various observational techniques, such as fluorescence spectroscopy, radio-frequency transmission spectroscopy and direct ion signal measurement using vapor cells with internal electrodes, to comprehensively characterize and determine the relevant parameters.  Preliminary results have shown delay in the ion signal compared to the optical transmission, but the complete results of such studies will be presented in future publications. Our findings enhance the understanding of nonequilibrium dynamics in hot Rydberg atomic gases, offering valuable insights for studies in many-body physics and nonlinear phenomena.

\begin{backmatter}
\bmsection{Funding}
This work is supported by the Innovation Program for Quantum Science and Technology under Grant No.2023ZD0300900, National Natural Science Foundation of China under Grant No.12027806, Fund for Shanxi ``1331 Project'', and Hanjiang National Laboratory.

\bmsection{Acknowledgments}

We are grateful to Wei Yi and Yu Chen for fruitful discussions. 

\bmsection{Disclosures}
The authors declare no conflicts of interest.

\bmsection{Data Availability Statement}
Data underlying the results presented in this paper are not publicly available at this time but may be obtained from the authors upon reasonable request.

\end{backmatter}

\bibliography{refV2}

\end{document}